\documentclass[12pt]{article}
\usepackage{amssymb}
\usepackage[dvips]{epsfig}
\usepackage[utf8]{inputenc}
\usepackage{graphicx}
\usepackage{amsmath} 
\usepackage{bbm}
\usepackage{bm}

\usepackage{array}

\usepackage{hyperref}
\usepackage{longtable}

\newcommand{\s}{\,\,\,\,}

\def \lbdd {\overline{\Lambda}^{\ddagger}}

\newcommand{\dlb}{\delta_{\bar{\Lambda}}}

\newcommand{\be}{\begin{equation}}
\newcommand{\ee}{\end{equation}}
\def\bea{\begin{eqnarray}}
\def\eea{\end{eqnarray}}
\newcommand{\bn}{\begin{eqnarray}}
\newcommand{\en}{\end{eqnarray}}
\newcommand{\nn}{\nonumber}
\newcommand{\no}{\noindent}

\newcommand{\p}{\partial}
\def\bea{\begin{eqnarray}}
\def\eea{\end{eqnarray}}
\newcommand{\beq}{\begin{eqnarray}}
\newcommand{\eeq}{\end{eqnarray}}

\def \osigma {\overline \sigma}

\def \olamb {\overline \Lambda }
\def \olambt {\overline \Lambda^T }

\begin{document}

\title{\textbf{Unitarity of partially broken massless higher-spin models}}

	\author{ D. Dalmazi\footnote{denis.dalmazi@unesp.br} , B. dos S. Martins\footnote{bruno.s.martins@unesp.br} , E. L. Mendonça\footnote{elias.leite@unesp.br} , R. Schimidt Bittencourt\footnote{raphael.schimidt@unesp.br} . \\
		\textit{Departamento de Física, FEG-UNESP}\\
        \textit{Av. Dr. Ariberto Pereira da Cunha, 333, Guaratinguetá, SP, Brazil.}}
	\date{\today}
	\maketitle

\begin{abstract}

Here we suggest a partially broken version of the Skvortsov-Vasiliev (SV) model for massless particles of arbitrary integer spin $s\ge 3$. The traceless gauge parameter of the Weyl transformation is now  required to be transverse. In the light-cone gauge we offer a simple proof that the model contains only spin-$s$ helicities as propagating modes. In the $s=3$ and $s=4$ cases we are able to calculate the two point amplitude and confirm unitarity in a explicitly Lorentz invariant way. 

For the recently suggested  partially broken Fronsdal model, for $s=3$ and $s=4$, unitarity is also confirmed in a Lorentz invariant way from their two point amplitudes. In both Fronsdal and SV partially broken models the two point amplitudes differ from their unbroken counterparts by contact terms which guarantees the same particle content but indicates potential non-equivalence of possible interacting terms.

\end{abstract}

\newpage

\section{Introduction}

In the work \cite{fp} published in 1939, Fierz and Pauli (FP) developed a Lorentz covariant description of massive spin-2 particles in terms of a symmetric rank-2 tensor. 
The massive action of \cite{fp} was generalized to arbitrary integer spin in 1974 by Singh and Hagen \cite{sh}. In the massless limit of \cite{sh} we are left with two symmetric traceless fields, one of rank-$s$ and another one of rank-$(s-2)$ which can be combined into one rank-$s$ double traceless field. This is how the Fronsdal model for massless particles of arbitrary integer spins was born in 1978 \cite{fronsdal}.
The Fronsdal model is invariant under higher spin (HS) analogues of spin-2 linearized diffeomorphisms\footnote{We work in flat Minkowski spacetime with $\eta = \mathrm{diag}(-,+,+,\dots,+)$ and in $D \geq 3$ dimensions. 
	Throughout the main text, symmetrizations have no weights, 
	for instance, $T_{(\mu\nu)} = T_{\mu\nu} + T_{\nu\mu}$, 
	while in the appendix the spin-projector definitions require weighted symmetrizations. In part of the work we adopt the compact notation of \cite{fs1}, where symmetrizations are implicit. For example, $\partial^k h = \partial_{(\mu_1}\cdots \partial_{\mu_k} h_{\mu_{k+1} \dots \mu_{k+s})}$, $
	\eta^k h = \eta_{(\alpha_1\beta_1}\cdots \eta_{\alpha_k\beta_k} h_{\mu_{1} \dots \mu_{s})}
$.  Dots indicate contractions: $(\partial \cdot)^k h = \partial^{\mu_1} \cdots \partial^{\mu_k}
	h_{\mu_1 \dots \mu_k\mu_{k+1}\cdots\mu_s}$, $(\eta \cdot)^k h = \eta^{\mu_1 \mu_2} \cdots \eta^{\mu_{2k-1}\mu_{2k}} 
	h_{\mu_1\cdots\mu_{2k-1}\mu_{2k}\mu_{2k+1} \cdots \mu_s} \equiv h^{[k]}$. Primes denote lower $k$ traces, e.g. $h'=\eta\cdot h$ and $h''=h^{[2]}$. A single bar denotes a traceless object 
	($\eta^{\mu_1\mu_2} \bar{h}_{\mu_1\mu_2 \dots \mu_s} = 0$). The indices 
 $``T"$ and $\ddagger$ indicate vanishing single and double divergence respectively, for instance, $\p\cdot u^T=0=\p\cdot\p\cdot v^{\ddagger}$.
}: $\delta_F h = \p\, \overline{\Lambda} $. The gauge parameter $\olamb$ is a rank-$(s-1)$ totally symmetric traceless tensor. For review works on higher spin particles see
\cite{vasiliev_r,st_r,rt,snow,book,vasilievr2}.

In the $s=2$ case the Fronsdal model becomes  the linearized version of the Einstein-Hilbert (EH) general relativity, invariant under $\delta h_{\mu\nu} = \p_{\mu}\Lambda_{\nu} + \p_{\nu}\Lambda_{\mu}$.  There is however, an alternative description of massless spin-2 particles in terms of a rank-2 symmetric tensor, see \cite{blas}, it is invariant under Weyl and transverse diffeomorphisms (WTDiff),  $\delta h_{\mu\nu} = \p_{\mu}\Lambda_{\nu}^T + \p_{\nu}\Lambda_{\mu}^T + \eta_{\mu\nu}\, \sigma $. The WTDiff model is the linearized version of unimodular gravity. In 2008 Skvortsov and Vasiliev (SV) generalized the spin-2 WTDiff model to arbitrary integer spins $s\ge 2$ \cite{sv}. The spin-$s$ SV model is described in terms of a double traceless totally symmetric rank-$s$ field. It is invariant under HS traceless and transverse diffeomorphisms and traceless Weyl transformations:  $\delta_{SV} h = \p \olambt + \eta\, \osigma $. The number of independent gauge parameters of $\delta_{SV} h$ and $\delta_{F} h$ is the same. 

One has recently found \cite{pbf} a partially broken version of the Fronsdal model (PBF)  for arbitrary integer spin. It is invariant under traceless HS diffeomorphisms further restricted by a vanishing double divergence condition on the gauge parameter: $\delta_{PBF} h = \p \lbdd$, where the restriction $\p\cdot\p\cdot \lbdd =0$  decreases the symmetry and allows for more general source couplings to the double traceless higher spin field when compared to the Fronsdal model. It is natural to ask whether we could also have a partially broken version of the SV model, PBSV henceforth. Indeed, the main purpose of the present work is to show that this is a viable model for massless bosonic higher spin particles.

 In section 2, using a basis of spin projection operators on rank-3 and rank-4 symmetric tensors, explicitly displayed in the appendix, we define the spin-3 and spin-4 PBF and PBSV models in a unified way and present a conjecture on the meaning of some determinants appearing in the two point amplitude obtained via projection operators for arbitrary integer spin. The PBSV model is presented in detail in section 3  for arbitrary integer spin in terms of a double traceless  totally symmetric rank-$s$ field. We prove that it shares the same particle content as the SV model, although it possesses less gauge symmetry. It is invariant under traceless and transverse HS diffeomorphisms and traceless and transverse Weyl transformations: $ \delta_{PBSV} h = \partial \bar{\Lambda}^T + \eta\, \bar{\sigma}^T $. The new restriction $\p\cdot \bar{\sigma}^T =0 $ is a rank-$(s-3)$ traceless symmetric tensor just like $\p\cdot\p\cdot \lbdd =0$. 

In \cite{pbf} it has been shown
that the PBF model for arbitrary integer spin-$s$ only contains massless spin-$s$ propagating modes without lower spin ghosts. For the spin-3 case the proof is gauge invariant, though not explicitly Lorentz invariant. For higher spins the demonstration is carried out in the light cone gauge. In subsections 4.1 and 4.2 we calculate, respectively, the two point amplitude for spin-3 and spin-4 PBF models in a explicitly Lorentz covariant way and compare the results with the Fronsdal theory, thus confirming the unitarity of the new models.
Likewise, in subsection 4.3 and 4.4 we calculate the two point amplitude for the PBSV model explicitly for $s=3$ and $s=4$ respectively. The amplitudes in both cases (PBSV and PBF) differ from the Fronsdal results by contact terms. In section 5 we draw our conclusions. Section 7 contains an Appendix with higher spin projection operators acting on rank-3 and rank-4 totally symmetric fields. 

\section{Irreducible spin-3 and spin-4 models}

\subsection{The spin-3 case}

Recently, we have shown in \cite{tdiffspin3} that the most general second-order, local, massless spin-3 model that is free of spin-2 ghosts, described by a symmetric rank-3 tensor field $h_{\mu\nu\rho} = h_{(\mu\nu\rho)}$ is governed by the following action, which depends on three arbitrary parameters $(a, b, c)$,
\be 
S(a,b,c) = \int d^Dx\, \left\lbrack \frac 12 h\, \Box \, h 
+ \frac 32 (\p\cdot h)^2 
+ 3\, a \, h' \p\cdot\p\cdot h 
- \frac 32\, b \, h'\, \Box \, h' 
+ \frac 34\, c \, (\p\cdot h')^2 \right\rbrack . \\ 
\label{sabc} 
\ee

\noindent One can verify that the action $S(a,b,c)$ is invariant under traceless and transverse HS diffeomorphisms, explicitly,
\be 
\dlb h_{\mu\nu\rho} = \p_{(\mu}\bar{\Lambda}^T_{\nu\rho)} 
\qquad ; \qquad  
\p^{\nu}\bar{\Lambda}^T_{\nu\rho} = 0 = \eta^{\nu\rho}\bar{\Lambda}^T_{\nu\rho} \, . 
\label{difftt} 
\ee 

In terms of the spin-s operators $P_{ij}^{(s)}$ (see the appendix) acting on symmetric rank-3 fields, the Lagrangian can be expressed as:
\begin{align}
	\mathcal{L} &=h_{\mu\nu\alpha}\left[
	\frac{\Box}{2} P_{11}^{(3)} 
	+ \frac{\Box}{2} \left[ 1 - b(D+1) \right] P_{11}^{(1)} 
	+ \frac{\Box}{2} (2a - b - 1) P_{22}^{(1)} \right.\nn \\
	&\quad + \frac{\Box}{2} \sqrt{D+1} (a - b) \left(P_{12}^{(1)} + P_{21}^{(1)}\right) 
	- \frac{(D-1)}{4} \, \Box (2b + c) P_{11}^{(0)}\nn \\
	&\quad \left.
	- \frac{\Box}{4} (-12a + 6b + 3c + 4) P_{22}^{(0)}  
	- \frac{\Box}{4} \sqrt{3(D-1)} (-2a + 2b + c) \left(P_{12}^{(0)} + P_{21}^{(0)}\right)
	\right]^{\mu\nu\alpha}_{\beta\gamma\lambda}  h^{\beta\gamma\lambda} .
	\label{sand}\end{align}
Notice that the total sandwiched operator can be cast in a block-diagonal form:
\be
\mathcal{O} \;=\; \mathcal{O}^{(3)} \;\oplus\; \mathcal{O}^{(2)} \;\oplus\; \mathcal{O}^{(1)} \;\oplus\; \mathcal{O}^{(0)} \,,
\ee
where each block $\mathcal{O}^{(s)}$ corresponds to the contribution of the corresponding spin-$s$ sector. 
The precise size of each block depends on the number of independent projectors in that sector.
In particular, the spin-0 and spin-1 sectors are described by $2\times 2$ matrices. Notice also that the absence of the spin-2 sector ($\mathcal{O}^{(2)}=0$), indicates precisely the presence of the rank-2 symmetry (\ref{difftt}). In the spin-0 and 1 sectors, we have from (\ref{sand}),
\be
\mathcal{O}^{(0)} \;=\; -
\begin{pmatrix}
	(D-1)(c+2\, b) & \sqrt{3(D-1)}(c+2\, b-2\, a) \\
	\sqrt{3(D-1)}(c+2\, b-2\, a) & (3\,c+4+6\, b-12\, a) 
\end{pmatrix}\, \frac {\Box}4, \ee

\be 
\mathcal{O}^{(1)} \;=\;
\begin{pmatrix}
	1-b\, (D+1) & \sqrt{D+1} (a-b) \\
	\sqrt{D+1} (a-b) & 2\, a - b-1
\end{pmatrix} \, \frac {\Box}2.\ee

The block structure of the total operator 
highlights how the determinants of these matrices control the invertibility 
of the kinetic operator necessary to obtain the propagator . Explicitly, the determinants are denoted by ${\cal K}^{(s)}_3$, where the upper index indicates the spin sector while the lower one represents the spin of the propagating modes, 
\be 
\mathcal{K}_3^{(0)}  \;=\; -\,\frac{\Box^2}{4}\,
\Big[\, (D-1)\,(3a^2 - 2b - c) \,\Big],
\qquad 
\mathcal{K}_3^{(1)}  \;=\; -\,\frac{\Box^2}{4}\,
\Big[\, D\,(a^2 - b) + (a-1)^2 \,\Big]. \label{det3}
\ee

Up to trivial local field redefinitions that we call $r$-shifts, $
h_{\mu\nu\alpha} \to h_{\mu\nu\alpha} +r\,\eta_{(\mu\nu}h^{'}_{\alpha)} $, there are four possible sets of choices of parameters $(a,b,c)$ which renders (\ref{sabc}) an irreducible description of massless spin-3 particles. Namely, 

\bea  (\mathcal{K}_3^{(0)},\mathcal{K}_3^{(1)}) &=& (0,0) \quad ;  \quad a\ne  2/(D+2) \qquad ({\rm Fronsdal \,\, model}) \quad \label{f} \\
(\mathcal{K}_3^{(0)},\mathcal{K}_3^{(1)}) &=& (0,0) \quad ;  \quad a= 2/(D+2) \qquad ({\rm SV\,\, model}) \quad \label{sv} \\
\mathcal{K}_3^{(0)} \ne 0 \, {\rm and}\,\,  \mathcal{K}_3^{(1)} &=& 0 \qquad \quad \!;  \quad a\ne 2/(D+2) \qquad ({\rm PBF \,\, model}) \quad \label{nf} \\
\mathcal{K}_3^{(0)} \ne 0 \, {\rm and}\,\,  \mathcal{K}_3^{(1)} &=& 0 \qquad \quad \!;  \quad a=2/(D+2) \qquad ({\rm PBSV \,\, model}) \quad \label{nsv}
\eea

\no The partially broken  SV (PBSV) model will be formally introduced in section 3. After treating the spin-4 in the next subsection, we   comment in subsection 2.3 on the interpretation of the determinants (\ref{det3}) and their higher spin analogues. 

\subsection{The spin-4 case}

The spin-4 case turns out to be somewhat more intricate from a technical standpoint. Unlike the spin-3 case, the spin-4 field naturally allows for the possibility of being double traceless. However, when analysing the spectrum of the model, it proves more convenient to restore the double trace by means of the field redefinition: $h_{\mu\nu\alpha\beta} \to h_{\mu\nu\alpha\beta} - \eta_{(\mu\nu}\eta_{\alpha\beta)} h''/D(D+2)$, which automatically introduces a scalar Weyl symmetry $\delta h_{\mu\nu\alpha\beta}=\eta_{(\mu\nu}\eta_{\alpha\beta)}\sigma$. Applying this redefinition to the action $S(a,b,c)$ leads to a new form, $\tilde{S}(a,b,c)$, which is given, using the compact notation of \cite{fs1}, by:

\be
\begin{aligned}
	\tilde{S}(a,b,c) = \int d^D x \, \bigg\lbrack 
	&\frac{1}{2} h \Box h 
	+ 2 (\partial \cdot h)^2 
	+ 6a \, h'\, \partial \cdot \partial \cdot h  
	- 3b \, h' \Box h' \\
	&+ 3c \, (\partial \cdot h')^2 
	+ c_1 \, h'' \Box h'' 
	+ c_2 \, h'' \, \partial \cdot \partial \cdot h' 
	\bigg\rbrack
\end{aligned}\label{S4abc}
\ee

\no with:

\be c_1 = \frac{3\left[ 8a + 2D^2 b + D(4b - 2c-1) - 4(1 + c) \right]}{2 D^2 (D + 2)} \quad;\quad
c_2 = \frac{12 - 6(D + 4)a + 6(D + 2)c}{D(D + 2)}.\ee

In terms of the spin-4 projection operators defined in the appendix, the Lagrangian corresponding to the action (\ref{S4abc}) can be written as:

\begin{align}
	\mathcal{\tilde{L}} = h_{\mu\nu\rho\sigma} \Bigg[ 
	&\frac{\Box}{2} P_{11}^{(4)} 
	+ \frac{\Box}{2} \left( 1 - (D+3)b \right) P_{11}^{(2)} 
	+ \frac{\Box}{2} (2a - b - 1) P_{22}^{(2)} \nonumber \\
	&+ \frac{\Box}{2} \sqrt{D+3}(a - b) \left(P_{12}^{(2)} + P_{21}^{(2)}\right) 
	- \frac{\Box(D+1)}{4} (2b + c) P_{11}^{(1)} \nonumber \\
	&+ \frac{\Box}{4} (12a - 6b - 3c - 4) P_{22}^{(1)} 
	+ \frac{\Box}{4} \sqrt{3(D+1)} (2a - 2b - c) \left(P_{12}^{(1)} + P_{21}^{(1)}\right) \nonumber \\
	&+ \frac{\Box}{2} \left[ 1 - 2(D+1)b + \frac{2(D-1)(D+1)}{3} c_1 \right] P_{11}^{(0)} \nonumber \\
	&+ \frac{\Box}{2} \left[ 2a - b - 1 - (D-1)(b + c) + \frac{2(D-1)}{3} (2c_1 + c_2 ) \right] P_{22}^{(0)} \nonumber \\
	&+ \frac{3\Box}{2} \left[ 4a - 2b - 2c - 1 + \frac{2}{3}(c_1 + c_2) \right] P_{33}^{(0)} \nonumber \\
	&+ 9\sqrt{2(D+1)} \Box \left[ \frac{a - b}{18} + \frac{(D-1)}{108} (4 c_1 + c_2) \right] 
	\left(P_{12}^{(0)} + P_{21}^{(0)}\right) \nonumber \\
	&+ \frac{\sqrt{6(D-1)}}{2} \Box \left[ a - b - c + \frac{1}{6} (4c_1 + 3c_2) \right] 
	\left(P_{23}^{(0)} + P_{32}^{(0)}\right) \nonumber \\
	&+ \frac{\sqrt{3(D-1)(D+1)}}{6} \Box (2c_1 + c_2) 
	\left(P_{13}^{(0)} + P_{31}^{(0)}\right)
	\Bigg]^{\mu\nu\rho\sigma}_{\alpha\beta\gamma\lambda} 
	h^{\alpha\beta\gamma\lambda}
	\label{lagrangian_spin4_projectors}
\end{align}

Once again, we examine the determinants associated with the lower-spin sectors of the propagator in order to identify the possible gauge symmetries and the corresponding models. In this case, we obtain  
\[
\mathcal{K}_4^{(0)} = 0 \quad ; \quad 
\mathcal{K}_4^{(1)} = -\,\frac{\Box^2}{4}\,\Big[(D+1)(3a^2 - 2b - c)\Big]
\quad ; \quad
\mathcal{K}_4^{(2)} = -\,\frac{\Box^2}{4}\,\Big[(D+2)(a^2 - b) + (a-1)^2\Big].
\]

The first point to notice is that $\mathcal{K}_4^{(0)}=0$, which makes the spin-0 sector singular. This is a consequence of the scalar Weyl symmetry $\delta h = \eta\,\eta \,\sigma$ and require an appropriate gauge fixing in order to obtain the propagator. Regarding the spin-2 and spin-1 sectors, it is completely analogous to the spin-1 and spin-0 sectors of the spin-3 case of the previous subsection. Namely, we have four irreducible ghost free models for massless spin-4 particles according to:

\bea  (\mathcal{K}_4^{(1)},\mathcal{K}_4^{(2)}) &=& (0,0) \quad ;  \quad a\ne  2/(D+4) \qquad ({\rm Fronsdal \,\, model}) \quad \label{f4} \\
(\mathcal{K}_4^{(1)},\mathcal{K}_4^{(2)}) &=& (0,0) \quad ;  \quad a= 2/(D+4) \qquad ({\rm SV\,\, model}) \quad \label{sv4} \\
\mathcal{K}_4^{(1)} \ne 0 \, {\rm and}\,\,  \mathcal{K}_4^{(2)} &=& 0 \qquad \quad \!;  \quad a\ne 2/(D+4) \qquad ({\rm PBF \,\, model}) \quad \label{nf4} \\
\mathcal{K}_4^{(1)} \ne 0 \, {\rm and}\,\,  \mathcal{K}_4^{(2)} &=& 0 \qquad \quad \!;  \quad a=2/(D+4) \qquad ({\rm PBSV \,\, model}) \quad \label{nsv4}
\eea

\no Next we conjecture on the form and interpretation of the determinants $\mathcal{K}_s^{(j)}$ with $j=s-2$ and $j=s-3$ for arbitrary values of the spin-$s$.

\subsection{A conjecture}


As shown in \cite{pbf}, starting with the following three parameters Lagrangian given in terms of a double traceless rank-$s$ symmetric field,

\be  \mathcal{L}^{(s)}(a,b,c)  = \frac{1}{2}h\Box h + \frac{s}{2}(\partial \cdot h)^2 +  \frac{a\,s(s-1)}{2} \partial \cdot \partial \cdot h \ h' +  \frac{b\, s(s-1)}{4}(\partial h')^2 + \frac{3\,c}4\,\binom{s}{3} (\partial \cdot h')^2 \, , \label{labc} \ee

\no after performing the following invertible field redefinition\footnote{Assuming $a\ne a_s^*\equiv 2/[D+2\, (s -2)]$.}, which preserves the double traceless nature of the field,

\be \label{transfqualquer} h_{\mu_1 ... \mu_s} \rightarrow h_{\mu_1 ... \mu_s} + r\, \eta_{(\mu_1 \mu_2} h^{\alpha}_{\phantom \alpha \alpha \mu_3 ... \mu_{s-2})} \quad , \quad  r = \frac{1-a}{a[D+2\, (s -2)]-2} \ee

\no the Lagrangian (\ref{labc}) splits into the sum of the usual Fronsdal Lagrangian $\mathcal{L}^{(s)}_F=\mathcal{L}^{(s)}(1,1,1)$ and two trace dependent terms. Namely, 

\be \mathcal{L}^{(s)}(a,b,c) = {\cal L}_F^{(s)}(h) + \frac{s(s-1)(D+2s-6)}{4[a(D+2\, s -4)-2]^2}\left\lbrack f_b^{(s)} \, h^{'}\, \Box \, h^{'} + \frac{f_c^{(s)}(s-2)}2 \left( \p\cdot h^{'}\right)^2 \right\rbrack ,\label{BC} \ee

\no where 

\be  
f_b^{(s)} = (D+2s-6)(a^2-b) +(a-1)^2 \quad , \quad 
f_c^{(s)} = (D+2s -6)(c-a^2)+2\, (a-1)^2  . \label{fbfc} \ee

In \cite{pbf} one has argued that the content of (\ref{BC}) can be read off
from the Fronsdal theory and the trace dependent terms independently as if the trace $h'$ was an independent field, though the field $h$ in ${\cal L}_F^{(s)}(h)$ is not traceless. 

The Fronsdal theory  is an irreducible theory containing only  spin-$s$ modes. The spin-$(s-1)$ modes are not present in the whole Lagrangian (\ref{BC}) due to the invariance under $\delta h = \p \, \overline{\Lambda}^T$ which holds for arbitrary values of the triple $(a,b,c)$, recall that $\overline{\Lambda}^T$ stems from the action of a spin-$(s-1)$ projection operator on an arbitrary rank-$(s-1)$ tensor. Since $( \p\cdot h^{'})$ is a rank-$(s-3)$ tensor, the spin-$(s-2)$ modes can only be present in the $h'\, \Box \, h'$ term, which explains why $\mathcal{K} _s^{(s-2)}$ are proportional to $f_b^{(s)}$ for $s=3$ and $s=4$. We conjecture that $\mathcal{K} _s^{(s-2)}$ is proportional to $f_b^{(s)}$ for arbitrary integer spin-$s$. At $f_b^{(s)}=0$ the symmetry increases:
$\overline{\Lambda}^T \to \lbdd$. The rank-$(s-2)$ fields present in $\p \cdot \lbdd$ generate the symmetry responsible for  $\mathcal{K} _s^{(s-2)}=0$. The condition $f_b^{(s)}=0$ defines the PBF model of \cite{pbf}. There is no need of setting $f_c^{(s)}=0$ to get rid of the spin-$(s-3)$ modes present in $( \p\cdot h^{'})^2$ because this term has no particle content by itself, similarly in the PBSV model to be defined in section 6. Moreover, the spin-$j$ modes with $0\le j\le (s-4)$ are eliminated by the double traceless condition.  At the ``Fronsdal point''  $f_b^{(s)}=0=f_c^{(s)}$
we have $c=  c_S \equiv 3\, a^2 - 2\, b$, the symmetry increases again $\lbdd \to \overline{\Lambda}$  and the rank-$(s-3)$ fields present in $\p\cdot\p\cdot \overline{\Lambda} $ generate the symmetry responsible for $\mathcal{K} _s^{(s-3)}=0$. We conjecture that $\mathcal{K} _s^{(s-3)}$ is always proportional to $(c-3\, a^2 + 2\, b)$ for any integer spin-$s$. Notice that even at $f_b^{(s)}\ne 0$, the condition
$f_c^{(s)}=2\,f_b^{(s)}\ne 0$ leads to $c=c_S\equiv 3\, a^2 - 2\, b$ indicating a
rank-$(s-3)$ symmetry without a rank-$(s-2)$ one. Indeed, at that specific point the two trace dependent terms in (\ref{BC}) become a spin-$(s-2)$ SV model \cite{sv} describing only spin-$(s-2)$ propagating modes. A simpler example is the $s=3$ case where there appears the Maxwell theory where the spin-0 modes present in $f_b\,  A_{\mu}\Box A^{\mu} + f_c (\p\cdot A)^2/2$ disappear at $f_c=2\, f_b$ due to the spin-0 $U(1)$ gauge symmetry and we are left only with spin-1 modes.

It is worth emphasizing that the main obstacle that prevents a proof of the conjectured factorization of $(\mathcal{K} _s^{(s-2)}, \mathcal{K} _s^{(s-3)})$ in terms of  $(f_b^{(s)},c-3\, a^2 +2\, b)$ is, besides of course the absence of an explicit calculation of those determinants for arbitrary integer spin-s, the fact that we have no rigorous proof for arbitrary integer spin-s that, see \cite{pbf},  the set of invariants under  $\delta h = \p\,\lbdd $ can be decomposed in a invertible way in terms of helicity variables. In particular, we do not know the analogue of the spin-3 matrix (3.17) of \cite{pbf} for arbitrary integer spin-s, so we are not able to check that the matrix determinant is non vanishing. Consequently, we have no rigorous proof for $s\ge 4$ that the particle content of (\ref{BC}) can be read off independently from the Fronsdal action and from the trace dependent terms inside brackets.

\section{Partially broken Skvortsov-Vasiliev model}

\subsection{Definition and gauge symmetries}

In this section we introduce the ``partially broken Skvortsov-Vasiliev" (PBSV) model for arbitrary integer spin-$s$. It is a one parameter model  given by a Lagrangian $\mathcal{L}_{PBSV}^{(s)}(c)$  formulated in terms of a double traceless ($h''=0$) rank-$s$ symmetric field $h_{\mu_1\mu_2\cdots \mu_s}$. Namely,

\bea \mathcal{L}_{PBSV}^{(s)}(c) \!\!\! &=& \!\!\!\frac{1}{2}h\Box h + \frac{s}{2}(\partial \cdot h)^2 +  \frac{a_s^*\,s(s-1)}{2} \partial \cdot \partial \cdot h \ h' +  \frac{b_s^*\, s(s-1)}{4}(\partial h')^2 + \frac{3\,c_s^*}4\,\binom{s}{3} (\partial \cdot h')^2 \nn\\ &+&\!\!\!  \frac{3\,(c-c_s^*)}4\,\binom{s}{3} (\partial \cdot h')^2\label{lpbsv}\quad ,\eea
where

\be a_s^* = \frac{2}{D+2(s-2)}, \quad b_s^* = \frac{D+2(s-1)}{[D+2(s-2)]^2}, \quad  c_s^* = -2\,\frac{D+2(s-4)}{[D+2(s-2)]^2}, \qquad c\ne c_s^* \label{star}\ee

 The case $c=c_s^*$ is exactly the SV theory \cite{sv}
which is invariant under traceless and transverse diffeomorphisms ($\overline{\Lambda}^T$) and traceless Weyl transformations ($\overline{\sigma}$) which amounts to  $\delta_{SV}h = \partial \overline{\Lambda}^T  + \eta \, \overline{\sigma}$.  Since the field is double traceless, we can use the traceless Weyl symmetry and gauge away the trace completely ($h'=\eta \cdot h =0$) and reformulate the SV model in terms of a traceless field ($\overline{h}$)  with the symmetry $\delta \, \overline{h} = \partial \,\overline{\Lambda}^T $, as in \cite{sv}. The last term in (\ref{lpbsv}) breaks the traceless Weyl symmetry to its transverse version. The full symmetry of (\ref{lpbsv}) becomes
\be \label{deltansv} \delta h = \partial \bar{\Lambda}^T + \eta \bar{\sigma}^T \quad . \ee

\subsection{Equations of motion, constraints and degrees of freedom}

In the PBSV model the reduced Weyl symmetry only allow us to gauge away the transverse part of the field's trace. However, as we will see, a new gauge-independent constraint arises which eliminates the longitudinal components of the trace on shell. The equations of motion from (\ref{lpbsv}) are given by the vanishing double-traceless Euler tensor:
\be \label{eompbsv}E \equiv \Box h - \p \ \p\cdot h + a_s^* \eta \ \p \cdot \p \cdot h + a_s^* \p \p h' - b_s^*\Box\eta h' - \frac{c}{2}\eta \ \p \ \p \cdot h' - \Upsilon \ \eta \ \eta \ \p \cdot \p \cdot h' = 0\ee
where $\Upsilon = -2c/[D+2(s-4)]$  assures that the Euler tensor is identically double-traceless. 

We then have the trace of (\ref{eompbsv}),
\be \label{trace_euler_pbsv}E' = \{ 2a_s^* - 1 - \frac{c}{2}[D+2(s-2)] \} \,\p \,\p \cdot h' + \{ a_s^* - c - \Upsilon[D+2(s-3)] \}\, \eta \,\partial \cdot \partial \cdot h'. \ee
The reader can check that if $c=c_s^*$ we have that $E'\equiv0$ holds identically due to the full traceless Weyl symmetry.

For arbitrary functions $\psi$ and $\chi$ of fields that vanish at infinity,  equations of the type $\p \psi + \eta \chi =0$ lead to $\psi = 0 = \chi$. Thus, from (\ref{trace_euler_pbsv}) we have the constraint
\be V_1 \equiv \label{vinculopbsv}\partial \cdot h'=0. \ee
The divergence of the equations of motion (\ref{eompbsv}) leads to
\be \partial \cdot E = \partial [(a_s^* -1)\partial\cdot \partial \cdot h + (a_s^*-b_s^*)\Box h'] + \eta a_s^* \ \partial \cdot \partial \cdot \partial \cdot h = 0 \ee
and then we have
\be \label{vinculosv} V_2 \equiv (a_s^* -1)\partial\cdot \partial \cdot h + (a_s^*-b_s^*)\Box h' = 0. \ee
The constraint (\ref{vinculosv}) is the same one which appears in the usual SV theory.

In order to fix the transverse and traceless Weyl symmetry ($\bar{\sigma}^T $) we can impose the gauge condition $h'=0$ which is  transverse and traceless  due to (\ref{vinculopbsv}) and $h''=0$ respectively. We can fix the symmetry  ($\overline{\Lambda}^T$) by further imposing $\partial\cdot h=0$, which is  traceless and transverse due to (\ref{vinculopbsv}) and  (\ref{vinculosv}) altogether with $h'=0$. Plugging all the constraints and gauge fixings in (\ref{eompbsv}) we obtain the expected wave equation for all field components: $\Box h=0$. Our gauge conditions are still invariant under the residual gauge symmetry $\delta h = \partial \bar{\Lambda}^T$ with $\Box \bar{\Lambda}^T=0$. So we can make the degrees of freedom count as follows.

 We have the double traceless symmetric field $h$, whose number of independent components $n(h)$ is given by the binomials difference:
\be n(h) = \binom{D+s-1}{D-1} - \binom{D+s-5}{D-1}.  \ee
The symmetric rank-$(s-1)$ traceless transverse diffeomorphism parameters have
\be n(\overline{\Lambda}^T) = \binom{D+s-2}{D-1} - \binom{D+s-3}{D-1} - \left[ \binom{D+s-4}{D-1} - \binom{D+s-5}{D-1} \right] \ee
independent components, and the rank-$(s-2)$ traceless transverse Weyl parameters have
\be n(\overline{\sigma}^T) = \binom{D+s-3}{D-1} - \binom{D+s-4}{D-1} - \left[ \binom{D+s-5}{D-1} - \binom{D+s-6}{D-1} \right] \ee
independent components. The constraint (\ref{vinculopbsv}) is given by a traceless rank-$(s-3)$ symmetric tensor, and then
\be n(V_1) = \binom{D+s-4}{D-1} - \binom{D+s-6}{D-1}. \ee
The constraint (\ref{vinculosv}) is represented by a rank-$(s-2)$ tensor, and its trace is identically null due to (\ref{vinculopbsv}), therefore
\be n(V_2) = \binom{D+s-3}{D-1} - \binom{D+s-5}{D-1} \ee
The number of propagating degrees of freedom can be obtained by the expression

 \be n_{\text{dof}} = n(h) - 2 \, n(\overline{\Lambda}^T) - n(\overline{\sigma}^T) - n(V_1) - n(V_2) \quad , \label{ndof}  \ee
 
\no  where the number of diffeomorphism parameters is multiplied by $2$ because of the residual gauge symmetry. Therefore we have
\be n_{\text{dof}} = \binom{D+s-1}{D-1} - \binom{D+s-5}{D-1} - 2\left[ \binom{D+s-2}{D-1} - \binom{D+s-4}{D-1} \right]\quad , \ee
\no which is the same counting of the Fronsdal, SV and PBF models. In particular, in $D=4$ and $D=5$ we have $n_\text{dof} = 2$  and $2\, s+1$ respectively, as expected. 

In the following we will be using the light cone gauge, which leaves no residual gauge symmetry, to check how the constraint (\ref{vinculopbsv}) compensates the reduced Weyl symmetry in order to eliminate the longitudinal propagation modes.

\subsection{PBSV content in the light cone gauge}
We now analyse the equations of motion of the PBSV model (\ref{lpbsv}) in the momentum space in light-cone coordinates, where we have  $(\eta_{-+},\eta_{+-},\eta_{IJ})= (-1,-1,\delta_{IJ})$ and $\eta_{++}=\eta_{--}=\eta_{\pm J}=0$, with $I,J = 1,2,\cdots, D-2$. We will suppose $p_+ \neq 0$ and use the index notation $\left(+(n)-(l)J(m)\right)$ indicating that we have $n$ indices ``$+$", $l$ indices ``$-$" and that $J(m)=J_1J_2\cdots J_m$.

The key idea of this demonstration is to use the traceless and transverse Weyl parameter to gauge away the transverse part of the field trace $h'$ and the constraint ({\ref{vinculopbsv}}) to eliminate the longitudinal components of $h'$, thus showing that $h'=0$ on-shell. The equations of motion and the remaining gauge symmetry are then reduced to those of the usual SV formulation, namely
\be \label{usualsv} p^2 h - p\ p\cdot h = 0; \quad h'=0; \quad \delta h = -i p\bar{\Lambda}^T. \ee

Starting from (\ref{deltansv}) one notes that
\be \delta h'_{\mu(s-2)} = [D+2(s-2)] \bar{\sigma}^T_{\mu(s-2)} \quad .\ee

\no The transversality condition $p\cdot \bar{\sigma}^T=0$ leads to a constraint on the Weyl parameters with at least one ``minus" index, 
\be \label{pdotsigmazero} \bar{\sigma}^T_{-\mu(s-3)} = \frac{1}{p_+} (p_J \bar{\sigma}^T_{J \mu(s-3)} - p_- \bar{\sigma}^T_{+\mu(s-3)} ),\ee
while the traceless condition $\eta\cdot\bar{\sigma}^T=0$ leads to
\be \label{sigmaprimezero} \bar{\sigma}^T_{+-\mu(s-4)} = \frac{1}{2} \bar{\sigma}^T_{JJ\mu(s-4)}. \ee
Taking (\ref{sigmaprimezero}) in (\ref{pdotsigmazero}) with $\mu(s-3)=+\mu(s-4)$, we have
\be \bar{\sigma}^T_{JJ\mu(s-4)} = \frac{2}{p_+}(p_I \bar{\sigma}^T_{I+\mu(s-4)} - p_- \bar{\sigma}^T_{+(2)\mu(s-4)}). \ee
Thus the remaining Weyl parameters allow us to impose the gauge condition
\be \label{gaugecond} h'_{+(s-2-m)\bar{J}(m)}=0, \quad m \in \{ 0,1,2,\dots, s-2 \}\ee
where
\be \nn \bar{J}(m) \equiv J_1 J_2 \dots J_l \dots J_k \dots J_m \quad \text{with} \quad l \neq k \ \forall \ l,k \in \{1,2,\dots , m \}. \ee

\no The constraint (\ref{vinculopbsv}) reads
\be \label{constraintmomentum} h'_{-\mu(s-3)} = \frac{1}{p_+} (p_J h'_{J\mu(s-3)} - p_- h'_{+\mu(s-3)}). \ee
Taking (\ref{constraintmomentum}) with $\mu(s-3) = +(s-3-m)\bar{J}(m)$, both terms in the right-hand side vanishes due to the gauge condition (\ref{gaugecond}), yielding 
\be \label{inducao0} h'_{-+(s-3-m)\bar{J}(m)} = 0, \quad m \in \{ 0,1,2,\dots,s-3 \}. \ee
Again, taking (\ref{gaugecond}) with $\mu(s-3) = -+(s-4-m)\bar{J}(m)$, where the right-hand side will vanish due to (\ref{inducao0}), we have
\be h'_{-(2)+(s-4-m)\bar{J}(m)} = 0, \quad m\in\{ 0,1,2,\dots, s-4 \}. \ee
One can then proceed in this way and find, after a finite number of steps,
\be \label{inducaofim} h'_{+(s-2-l-m)-(l)\bar{J}(m)} = 0 ,\ee 
with $l\in\{0,1,2,\dots,s-2\}$, $m\in \{ 0,1,2,\dots,s-2-l \}$.

The double-traceless condition in the field $h$ gives
\be \label{doubletracelightcone}h'_{JJ\mu(s-4)} = 2h'_{+-\mu(s-4)} .\ee
For $l \in \{0,1,2,\dots,s-4\}$ and $m\in \{0,1,2,\dots,s-4-l \}$, the right-hand side of (\ref{doubletracelightcone}) with $\mu(s-4) = +(s-4-l-m)-(l)\bar{J}(m)$ vanishes due to (\ref{inducaofim}), leading to
\be h'_{JJ+(s-4-l-m)-(l)\bar{J}(m)} = 0,\ee
therefore we have $h'=0$. The constraint (\ref{vinculosv}) then leads to $p\cdot p\cdot h=0$. The remaining diff symmetry and the equations of motion (see (\ref{eompbsv})) are thus given by the system (\ref{usualsv}), whose light-cone gauge analysis done in \cite{sv} showed that $h$ describes a spin-$s$ irreducible representation of the little group $O(D-2)$.

\section{Lorentz invariant unitarity proof}

\subsection{Unitarity of the spin-3 PBF model}

\noindent We can set up in (\ref{sabc}) $a = b = 1$ while $c$ remains arbitrary, except that $c \neq 1$, the theory corresponds to the partially broken Fronsdal (PBF) model, as defined in \cite{pbf}. In this case, the Lagrangian becomes invariant under doubly transverse and traceless diffeomorphisms (trDiff$^\ddagger$ henceforth), i.e.,

\be
\delta h_{\mu\nu\alpha} = \partial_{(\mu} \bar{\Lambda}^{\ddagger}_{\nu\alpha)} 
\quad ; \quad 
\eta^{\mu\nu} \bar{\Lambda}^{\ddagger}_{\mu\nu} = 0 
= \partial^\mu \partial^\nu \bar{\Lambda}^{\ddagger}_{\mu\nu}, \label{gs3}
\ee

\noindent which defines a symmetry of $S(1,1,c\neq 1)$. The particle content of this model was previously analysed in \cite{pbf} for $s=3$ using Lorentz non covariant gauge-invariants (Bardeen variables) and for arbitrary $s$ via the light cone gauge. In this section, we present a fully Lorentz-invariant derivation of the spectrum of the spin-3 PBF model by explicitly computing its propagator. This is achieved by introducing appropriate gauge-fixing terms associated with the trDiff$^\ddagger$ symmetry and subsequently saturating the resulting propagator with suitable sources.

The gauge-fixing term we consider is given by:

\be
\mathcal{L}_{gf} = \frac{\lambda}{2} \, f_{\mu\nu}^2
\ee 

\noindent where the gauge-fixing function is defined as:
\be
f_{\mu\nu} \equiv \frac{D}{2}  \partial_{(\mu} (\partial \cdot \partial \cdot h)_{\nu)}  
- (D-1) \, \Box \, (\partial \cdot h)_{\mu\nu} 
+ \eta_{\mu\nu} \left( \Box \, \partial \cdot h' - \partial \cdot \partial \cdot \partial \cdot h \right)
- \partial_\mu \partial_\nu \p\cdot h' \, . \label{gc3}
\ee

\no The reader can check that $\p^{\mu}\p^{\nu}f_{\mu\nu}=0=\eta^{\mu\nu} f_{\mu\nu}$ holds identically in agreement with the features of the gauge parameter $\lbdd_{\mu\nu}$. By including this gauge-fixing term, one obtains the propagator of the PBF model, which is proportional to
\bea
G^{-1} &=& \frac{2}{\Box} P_{11}^{(3)} 
- \frac{6}{\lambda \, \Box^3 (D-1)^2} P_{11}^{(2)} 
- \frac{2}{D \, \Box} P_{11}^{(1)} 
- \frac{12}{\lambda (D-2)^2 \Box^3} P_{22}^{(1)} \nn\\
&&\quad - \frac{1}{\Box (c - 2)} \left[ 
\frac{(3c - 2)}{(D - 1)} P_{11}^{(0)} 
+ (c + 2) P_{22}^{(0)} 
- \frac{c \sqrt{3(D - 1)}}{(D - 1)} (P_{12}^{(0)} + P_{21}^{(0)})
\right]\, ,
\eea

\noindent where we have suppressed the Lorentz indices for brevity. We now proceed to saturate $G^{-1}$ with external sources $T_{\mu\nu\alpha}$, which, due to the gauge symmetry, must satisfy the constraint:
\be
\partial_\mu T^{\mu\nu\alpha} = \eta^{\nu\alpha} \, \Omega + \partial^\nu \partial^\alpha J\label{constrs3}
\ee

 \no with $\Omega$ and $J$ arbitrary scalar functions. This constraint  ensures compatibility with the gauge symmetry (\ref{gs3}), see formula (5.1) of \cite{pbf}. Taking (\ref{constrs3}) into account we have:

\be
\bar{T} \, G^{-1} \, T 
= \frac{2}{\Box} \left[ (T^{\mu\nu\lambda})^2 
- \frac{3}{D} (T^{\mu})^2 
- 6 \, \Omega J \right] 
+ \frac{3}{D(c - 1)} \left[ c(D - 2) + 2 \right] J^2
\ee

In order to compare with the Fronsdal theory, we introduce $\widetilde{T}^{\mu\nu\lambda} $ according to: $T^{\mu\nu\lambda} \equiv \widetilde{T}^{\mu\nu\lambda} 
+ \, \eta^{(\mu\nu} \partial^{\lambda)} J/2$, such that $T^{\mu} = \widetilde{T}^{\mu} 
+ (D+2) \, \partial^{\mu} J/2$. This implies that $\p_{\mu}\widetilde{T}^{\mu\nu\lambda}=\eta^{\nu\lambda}\widetilde{\Omega}$ with $\widetilde{\Omega}=\Omega-\Box J/2$. Consequently, we have

\be
\bar{T} \, G^{-1} \, T 
= \underbrace{\frac{2}{\Box} \left[ \left(\widetilde{T}^{\mu\nu\lambda}\right)^2 
	- \frac{3}{D} \left(\widetilde{T}^{\mu}\right)^2 \right] 
}_{\text{Fronsdal}}
+ \frac{3}{(c - 1)} J^2 \label{48}
\ee

The terms involving $\widetilde{T}^{\mu\nu\lambda}$ and $\widetilde{T}^\mu$ reproduce exactly  the Fronsdal result which is known to describe a massless spin-3 particle. The analytic extra term proportional to $J^2$ assures that we have no further contribution to the particle content. It does not change the large distance behaviour of the Fronsdal theory, it is called a contact term. The existence of contact terms when we compare amplitudes of equivalent (dual) free models indicates that possible maps among gauge invariants of the corresponding dual theories do not hold at action level where we have the product of fields at the same space-time point, see comments in \cite{renato2} and \cite{gracia}.

%

\subsection{Unitarity of the spin-4 PBF  model}

Let us now analyse the unitarity of the spin-4 PBF model by fixing $a=b=1$ in (\ref{S4abc})  while keeping $c\neq 1$ arbitrary. In this setup, the Lagrangian becomes invariant under a reduced set of diffeomorphisms as compared to the Fronsdal model. More precisely, it exhibits the spin-4 analogue of the trDiff$^\ddagger$ symmetry, given by:

\be
\delta_1 h_{\mu\nu\alpha\beta} = \partial_{(\mu} \bar{\Lambda}^{\ddagger}_{\nu\alpha\beta)} 
\quad ; \quad 
\eta^{\mu\nu} \bar{\Lambda}^{\ddagger}_{\mu\nu\alpha} = 0 
= \partial^\mu \partial^\nu \bar{\Lambda}^{\ddagger}_{\mu\nu\alpha},
\ee
plus a Weyl scalar symmetry due to the restoration of the double trace of the field, given by:
\be
\delta_2 h_{\mu\nu\alpha\beta} = \eta_{(\mu\nu}\eta_{\alpha\beta)}\sigma.\ee
Then, an additional gauge-fixing term is required. Accordingly, we have:
\be
\mathcal{L}_{gf_1} = \frac{1}{2\lambda_1} \, \tilde{f}_{\alpha\beta\gamma}^2
\equiv \frac{1}{2\lambda_1} \left[ 
(D+1) f_{\alpha\beta\gamma}
+ (D-1) g_{\alpha\beta\gamma} 
\right]^2\quad;\quad 
\mathcal{L}_{gf_2} 
= \frac{1}{2\lambda_2} \left(h''\right)^2	
\ee
\no where:
\begin{align}
	f_{\alpha\beta\gamma} &= \Box^3 \, (\partial \cdot h)_{\alpha\beta\gamma} 
	- \Box^2 \, \partial_{(\alpha} \, (\partial \cdot\p\cdot h)_{\beta\gamma)} 
	+ \Box \, \partial_{(\alpha} \partial_\beta \, (\partial\cdot\p\cdot\p\cdot h)_{\gamma)} 
	- \partial_\alpha \partial_\beta\p_{\gamma} \, \partial \cdot\p\cdot\p\cdot\p\cdot h \nonumber \\
	&\quad - \frac{1}{(D+1)} \left\lbrack 
	\Box \, \theta_{(\alpha\beta} \, \partial_{\gamma)} \left(\partial \cdot\p\cdot\p\cdot\p\cdot h 
	- \Box \, \p\cdot\p\cdot h' \right) 
	- \Box^2 \, \theta_{(\alpha\beta} \left(\p\cdot\p\cdot\p\cdot h 
	- \Box \, \partial \cdot h' \right)_{\gamma)} 
	\right\rbrack
\end{align}
\bea
g_{\alpha\beta\gamma} &=& \Box^2 \, \partial_{(\alpha}(\partial\cdot\p\cdot h)_{\beta\gamma)} 
- 2 \Box \, \partial_{(\alpha} \partial_\beta \,(\partial\cdot\p\cdot\p\cdot h)_{\gamma)} 
+ 3 \, \partial_\alpha \partial_\beta \partial_{\gamma} \, \partial\cdot\p\cdot\p\cdot\p\cdot h 
\nn\\
&+& \frac{1}{(D-1)} \Box \, \theta_{(\alpha\beta} \, \partial_{\gamma)} 
\left( \partial\cdot\p\cdot\p\cdot\p\cdot h - \Box \, \p\cdot\p\cdot h' \right)
\eea

\no The reader can check that the first gauge condition obeys the same constraints of the gauge parameter $\lbdd_{\alpha\beta\gamma} $, namely, $\p \cdot \p \cdot \tilde{f} = 0 = \eta \cdot  \tilde{f}$.

After rewriting the gauge fixing terms using the spin-4 projection operator basis, one can finally derive the propagator associated with (\ref{lagrangian_spin4_projectors}). To obtain physical information from it, we consider the saturated propagator between external sources $T_{\mu\nu\alpha\beta}$ that satisfy a constraint analogous to the one employed in the spin-3 case, namely: $
\partial_\mu T^{\mu\nu\alpha\beta} = \eta^{(\nu\alpha} \, \Omega^{\beta)} + \partial^{(\nu }\partial^\alpha J^{\beta)}
$, where $(\Omega_{\beta)},J_{\beta})$ are arbitrary vectors.
Under this constraint we have,

\bea
T G^{-1} T &=& \frac{2}{\Box} \left\lbrack
(T^{\mu\nu\alpha\beta})^2 
- \frac{6}{(D+2)} (T^{\mu\nu})^2 
- 24 J^\mu \Omega_\mu 
- \frac{6(D^2 - 6D - 8)}{D(D+2)} (\partial \cdot J)^2\right\rbrack \nonumber \\
&+& \frac{12(c D + 2)}{(c - 1)(D + 2)} (J^\mu)^2\label{tgt1} 		
\eea
Again, in order to compare the transition amplitude from the Fronsdal model we introduce $\widetilde{T}^{\mu\nu\alpha\beta} $ as defined by:
\be	T^{\mu\nu\alpha\beta} \equiv \widetilde{T}^{\mu\nu\alpha\beta} 
+ \frac{1}{2} \, \eta^{(\mu\nu} \, \partial^\alpha J^{\beta)} 
- \frac{2}{D} \, \eta^{(\mu\nu} \, \eta^{\alpha\beta)} \, (\partial \cdot J)\label{decot}\ee

\no which is such that it satisfies a Fronsdal-like constraint, namely, 

\be \p^{\mu} \widetilde{T}_{\mu\nu\alpha\beta} = \eta_{(\nu\alpha}\tilde{\Omega}_{\beta )} \quad ; \quad \tilde{\Omega}_{\beta} \equiv {\Omega}_{\beta}- \frac {\Box J_{\beta}}2 - \frac{(D-4)}{2\,D} \p_{\beta} \, \p \cdot J  \, . \label{sourcec4} \ee

\no Substituting back (\ref{decot}) in (\ref{tgt1}) we have:
\be T G^{-1} T = \underbrace{\frac{2}{\Box} \left[ 
\left( \widetilde{T}^{\mu\nu\alpha\beta} \right)^2 
- \frac{6}{(D+2)} \left( \widetilde{T}^{\mu\nu} \right)^2 
\right]}_{Fronsdal}
+ \frac{12}{(c-1)} \left( J^\mu \right)^2
\ee
One observes that the contributions involving $\widetilde{T}^{\mu\nu\alpha\beta}$ and $\widetilde{T}^{\mu\nu}$ precisely reproduce the structure of the spin-4 Fronsdal propagator, which correctly describes a massless spin-4 particle. The remaining analytic term, proportional to $J^{\mu}J_{\mu}$, remember $c\ne 1$, is again a contact term with no influence in the large distance physics. This result confirms in a explicitly Lorenz invariant way  that the spin-4 PBF model ($c\ne 1$) has the same particle content as the spin-4 Fronsdal theory ($c=1$) as shown in \cite{pbf} in the light cone gauge.

\subsection{Unitarity of the spin-3 PBSV model}

In order to explore the particle content of the PBSV model in a explicitly Lorentz covariant way and compare it with its unbroken counter part, we now compute the two-point amplitude for the simplest case $s=3$. This computation requires the introduction of a gauge-fixing term, since the Lagrangian (\ref{lpbsv}) is invariant under the gauge symmetry (\ref{deltansv}). The derivation of the propagator is made possible by the complete basis of spin-projectors acting on rank-3 symmetric tensors, which we present in full detail in the appendix.

We adopt a covariant gauge fixing scheme defined by the addition of the following term to the Lagrangian:
\be
\mathcal{L}_{gf} = \frac{\lambda_1}{2} \left( \bar{f}^T_{\mu\nu} \right)^2 + \frac{\lambda_2}{2} \left( g^T_\mu \right)^2,
\ee
where $\lambda_1$ and $\lambda_2$ are arbitrary gauge-fixing parameters, and the gauge conditions are given by
\be
\bar{f}_{\mu\nu}^{\,T} = \Box^2 \left( P^{(2)}_{ss} \right)_{\mu\nu}^{\ \ \, \alpha\beta} \left( \partial^\sigma h_{\alpha\beta\sigma} \right), \quad
g_\mu^{\,T} = \Box\, \theta_{\mu}^{\,\,\,\,\nu} \left( \partial^\alpha \partial^\beta h_{\alpha\beta\nu} \right).
\ee
Here, $P^{(2)}_{ss}$ denotes the transverse-traceless spin-2 projector  acting on symmetric rank-2 fields\footnote{\be \left( P_{SS}^{(2)} \right)^{\s\s\lambda\mu}_{\alpha\beta} = \frac 12 \left(
\theta_{\s\alpha}^{\lambda}\theta^{\mu}_{\s\beta} + \theta_{\s\alpha}^{\mu}\theta^{\lambda}_{\s\beta} \right) -
\frac{\theta^{\lambda\mu} \theta_{\alpha\beta}}{D-1}\ee}, while $\theta_{\mu}^{\,\,\,\,\nu}$ is given in (\ref{omegatheta}). Notice that $(\bar{f}_{\mu\nu}^{\,T},g_\mu^{\,T})$ have the same tensor properties of the gauge parameters
$(\overline{\Lambda}^T_{\mu\nu},\overline{\sigma}_{\mu}^T)$ respectively.

After implementing the gauge fixings, omitting indices for clarity, the full propagator is proportional to:
\bea
G^{-1} &=& \frac{2}{\Box} \, P_{11}^{\,3} 
+ \frac{2}{\lambda_1\, \Box^5} \, P_{11}^{\,2} 
+ \left\lbrack-\frac{2(D+2)^2}{D\,  \Box}+\frac{2(D+1)}{\lambda_2\Box^4}\right\rbrack P_{11}^{\,1} 
+ \frac{2}{\lambda_2\, \Box^4} \, P_{22}^{\,1}\nn \\
&+& \frac{2\sqrt{D+1}}{\lambda_2\, \Box^4} \left(P_{12}^{\,1} + P_{21}^{\,1}\right)
- \left[\frac{4(D-1)}{c(D+2)^2 + 2(D-2)} + \frac{3}{(D-1)}\right] \frac{1}{\Box} P_{11}^{\,0} \nn\\
&-& \left[ \frac{12}{c(D+2)^2 + 2(D-2)} + 1 \right] \frac{1}{\Box} P_{22}^{\,0}\nn\\
&+& \frac{3}{\sqrt{3(D-1)}\, \Box} \left[1 - \frac{4(D-1)}{c(D+2)^2 + 2(D-2)}\right] (P_{12}^{\,0} + P_{21}^{\,0}).\nn
\eea

\vspace{0.5em}

Next we saturate $G^{-1}$ it with external sources $T_{\mu\nu\alpha}$. Due to the gauge symmetry (\ref{deltansv}) the sources must satisfy the constraints
\be
\partial^{\alpha} T_{\alpha\mu\nu} = \eta_{\mu\nu} \, \Omega + \partial_{(\mu} J_{\nu)}, \qquad
T_{\mu} \equiv \eta^{\nu\lambda} T_{\mu\nu\lambda} = \partial_\mu \xi, \label{sources}
\ee
where $J_\mu$ and $\xi$ are arbitrary functions while	$\Omega = (\Box \xi - 2\,\partial \cdot J)/D$. Substituting the constrained source into the saturated amplitude, we obtain:
\bea
T G^{-1} T = \frac{2}{\Box} \left[ 
({T}^{\mu\nu\lambda})^2 + 6\, \xi\, \Omega 
- \frac{6(D+2)}{D} (J^\mu)^2\right] 
- \frac{6\left[c(D+2)^2 - 4\right]\, \xi^2}{D\left[c(D+2)^2 + 2(D-2)\right]}.
\eea

\vspace{0.5em}

In order to make contact with the Fronsdal theory, we introduce $\widetilde{T}_{\mu\nu\lambda}$ as follows,
\be
T^{\mu\nu\lambda} \equiv \widetilde{T}^{\mu\nu\lambda} + \eta^{(\mu\nu} J^{\lambda)},
\ee
where $\widetilde{T}^{\mu\nu\lambda}$ satisfies  Fronsdal type constraints 
$\p^{\mu}\widetilde{T}_{\mu\nu\lambda}=\eta_{\nu\lambda}\tilde{\Omega}$, where

\be \tilde{\Omega} = [ \Box\, \xi - (D+2) \p\cdot J]/D \quad . \label{omegatil} \ee

\no  Inserting the decomposition back into the amplitude, we arrive at the final expression:
\be
T\, G^{-1}\, T =\underbrace{\frac{2}{\Box} \left[ \left(\widetilde{T}^{\mu\nu\lambda}\right)^2 
	- \frac{3}{D} \left(\widetilde{T}^{\mu}\right)^2 \right] 
}_{\text{Fronsdal}}  
+ \frac{12}{(D+2)^2(c-c_3^*)}\, \xi^2.
\ee


\vspace{0.5em}

The structure of the result clearly shows that the spin-3 PBSV theory propagates the same physical degrees of freedom of the usual Fronsdal theory. The first term with a simple massless pole precisely reproduces the standard spin-3 amplitude. The second analytic term, proportional to $\xi^2$, does not propagate any degree of freedom. It is once again a contact term. Recall that we have assumed from the start that we are away from the $SV$ point
($c\ne c_3^*$), where $c_3^*=3\, (a_3^*)^2 - 2\, b_3^* = -2(D-2)/(D+2)^2$. The singularity at $c=c_3^*$ signals the appearance of a symmetry. At that point the transverse condition on the traceless Weyl gauge parameter is not required: $\overline{\sigma}^T \to \overline{\sigma}$.

 As we will see in the next subsection, the $s=4$ case exhibits similar features, with the gauge structure becoming a bit more rich as $s$ increases.
 
 \subsection{Unitarity of the spin-4 PBSV model}

In the spin-4 case, in order to analyse the PBSV model (\ref{lpbsv}), it is convenient to trade the double traceless field into an unconstrained one such that a scalar Weyl symmetry is introduced in order to keep the number of degrees of freedom, namely, 
$h_{\mu\nu\alpha\beta}\to h_{\mu\nu\alpha\beta}-\eta_{(\mu\nu}\eta_{\alpha\beta)}h''/D(D+2)$. In terms of the new unconstrained field the gauge symmetry transformations become 

\be 
\delta h_{\mu\nu\alpha\beta} = \partial_{(\mu} \bar{\Lambda}^{\,T}_{\nu\alpha\beta)} 
+ \eta_{(\mu\nu} \, \bar{\sigma}^{\,T}_{\alpha\beta)} 
+ \eta_{(\mu\nu} \eta_{\alpha\beta)}  \lambda \, .\label{gaugesym}
\ee

In order to compute the amplitude, we introduce a gauge-fixing term for each independent gauge symmetry appearing in (\ref{gaugesym}). Such that 

\be
\mathcal{L}_{gf} 
= \frac{\lambda_1}{2} \left( \bar{f}_{\mu\nu\alpha}^{T} \right)^2
+ \frac{\lambda_2}{2} \left( \bar{g}_{\mu\nu}^{T} \right)^2
+ \frac{\lambda_3}{2} \left( h^{\prime\prime} \right)^2,
\ee
where
\be 
\overline{f}_{\mu\nu\alpha}^{\,T}
= \Box^{3} \left( P_{11}^{(3)} \right)^{\gamma\rho\sigma}_{\mu\nu\alpha} \,
\partial^{\lambda} h_{\gamma\rho\sigma\lambda}\quad;\quad 
\overline{g}_{\mu\nu}^{\,T}
= \Box^{2} \left( P_{ss}^{(2)} \right)^{\alpha\beta}_{\mu\nu} \,
\partial^{\lambda} \partial^{\gamma} h_{\alpha\beta\gamma\lambda}.
\ee

The above operators $P_{11}^{(3)}$ (see appendix) and $ P_{ss}^{(2)}$ (see footnote 3) render  rank-3 and and rank-2  symmetric tensors respectively into transverse and traceless tensors as required by the symmetry (\ref{gaugesym}) which imply also the following constraints on the sources:
\be
\begin{aligned}
	\partial^\alpha T_{\alpha\beta\mu\nu} &= \eta_{(\beta\mu} \, \Omega_{\nu)} + \partial_{(\beta} J_{\mu\nu)} \\[6pt]
	T'_{\mu\nu} &= \eta_{\mu\nu}\,\psi + \partial_{\mu}\,\xi_{\nu}+\partial_{\nu}\,\xi_{\mu} \\[6pt]
	T'' &= 0
\end{aligned}
\ee
where we have introduced the ({\it a priori}) arbitrary  functions $\Omega_{\nu}$, $J_{\mu\nu}$, $\psi$, and $\xi_{\mu}$. The consistency of the constraints can be used to eliminate $\psi$  and $\Omega_{\nu}$ in favour of $\xi_{\nu}$ and $J_{\mu\nu}$, 
\be
	\psi = -\frac{2}{D}\,(\partial \cdot \xi) \quad , \quad 
	\Omega_{\mu} = \frac{1}{D+2} \left[ \frac{(D-2)}{D}\,\partial_{\mu}(\partial \cdot \xi) 
	+ \Box \xi_{\mu} - 2\,\partial^{\alpha} J_{\alpha\mu} - \partial_{\mu} J' \right].
\ee

The explicit form of $G^{-1}$ is technically involved. For simplicity, we present only the expression saturated by constrained sources, which takes the form
\begin{align}
	T\, G^{-1} T = \frac{2}{\Box} \Bigg[ &
	(T^{\mu\nu\lambda\rho})^2 
	- \frac{12(D+4)}{(D+2)} (J^{\mu\nu})^2 
	- \frac{12(D-2)}{D(D+2)} (\partial \cdot \xi)^2 \notag\\
	& - \frac{48}{D(D+2)}\, J'\partial \cdot \xi 
	+ \frac{12(D+4)}{D(D+2)} (J')^2 
	+ \frac{12\Box}{D+2} (\xi^\mu)^2- \frac{48}{(D+2)}\,  \xi_{\mu}\p_{\alpha}J^{\alpha\mu}  \Bigg] \notag\\
	& + \frac{48}{c(D+4)^2 + 2D}\, (\xi^\mu)^2. \label{tgt}
\end{align}

In order to compare with the spin-4 Fronsdal theory, we introduce $\tilde{T}_{\mu\nu\alpha\rho} $ according to:
\be
	T_{\mu\nu\alpha\rho} \equiv \tilde{T}_{\mu\nu\alpha\rho} 
	+   \eta_{(\mu\nu}J_{\alpha\rho)} - \frac{2}{D}\,\eta_{(\mu\nu}\eta_{\alpha\rho)}\,J' \quad . \label{ttil}	
\ee

\no The source $\tilde{T}_{\mu\nu\alpha\rho} $ satisfies Fronsdal type constraints:

\be \tilde{T}''=0 \quad , \quad \p^{\mu}\tilde{T}_{\mu\nu\alpha\rho}  = \eta_{(\mu\nu}\tilde{\Omega}_{\rho)} \quad , \quad \tilde{\Omega}_{\rho} \equiv \Omega_{\rho} - (\p\cdot J)_{\rho} + (2/D) \p_{\rho}\, J' \label{ttil2} \ee

\no After substituting (\ref{ttil}) in (\ref{tgt}) we have:

\be
T\, G^{-1} T = \underbrace{\frac{2}{\Box} \left[ 
\left( \widetilde{T}^{\mu\nu\alpha\beta} \right)^2 
- \frac{6}{(D+2)} \left( \widetilde{T}^{\mu\nu} \right)^2 
\right]}_{Fronsdal}
+ \frac{48}{(D+4)^{2}(c-c_4^*)}\,(\xi^{\mu})^{2},
\ee
which again correctly reproduces the transition amplitude of the Fronsdal theory, up to a non-propagating (contact) term. The singularity at $c=c_4^*=-2D/(D+4)^2$ stands for the appearance of a local symmetry. At $c=c_4^*$ we recover the SV model and the transverse (rank-2) Weyl symmetry is enlarged to full Weyl symmetry.

	\section{Conclusion}
    
    Here we have shown, complementing the recent results \cite{pbf} on the partially broken Fronsdal model, that a partially broken version of the Skvortsov-Vasiliev  model \cite{sv} can also be consistently formulated for arbitrary integer spin-$s$, see (\ref{lpbsv}). We have proved the consistency of the model in the light cone gauge in subsection 3.3 for arbitrary integer spin. The reduced gauge symmetry of the partially broken Fronsdal (PBF) model \cite{pbf} and the partially broken SV (PBSV) model allows for more general source couplings as compared to their unbroken counterparts.  The respective extra constraints on the gauge parameters, namely, $\p\cdot\p\cdot \lbdd=0$ and $\p\cdot \bar{\sigma}^T =0 $,  are both rank-$(s-3)$ traceless symmetric tensors. So both PBF and PBSV have the same number of gauge symmetries. In particular, for $s=3$ and $D=4$ we have 8 gauge parameters, see also \cite{bms3}, instead of the 9 parameters of the Fronsdal and SV models.
       
    In section 4 we have calculated the two point amplitude of the model (\ref{lpbsv})  and of the partially broken Fronsdal model suggested in \cite{pbf}  for the  spin-3 and spin-4 cases and shown that they differ from their unbroken versions by contact terms. In particular, for the spin-3 case, and $D=4$, our result for the partially broken SV model is in agreement with the recent result of \cite{bms3}. Namely\footnote{We thank the referee for suggesting a detailed comparison with \cite{bms3}.}, their table II contains two models containing only spin-3 physical  modes: ${\mathfrak{F}}_4$ and ${\mathfrak{E}}_6$, in their notation. The model ${\mathfrak{F}}_4$  is the  spin-3 version of the SV model \cite{sv}, since the first four equations in the second column of table II fix\footnote{The map between the Lagrangian (3) of \cite{bms3} and our Lagrangian (\ref{sabc}) is given by $(\kappa^{(4)}_1,\kappa^{(4)}_2,\kappa^{(4)}_4)= (3/4)(c,2\,b,-4\,a)$ and $(\kappa^{(4)}_3,\kappa^{(4)}_7)$=$(3,-1)/2$ while $\kappa^{(2)}_1=\kappa^{(2)}_2=0$.}   $(a,b,c)=(\frac 13,\frac 29,-\frac 19)$, in agreement with our formula (\ref{star}) at $D=4$, once we choose the overalll factor $\kappa^{(4)}_7=-\frac 12$.   The model ${\mathfrak{E}}_6$ is the spin-3 version of the PBSV model found here for arbitrary integer spin-s. In particular, as mentioned in \cite{bms3}, the model ${\mathfrak{E}}_6$  is the superposition of ${\mathfrak{F}}_4$ and the trivial (non propagating) model ${\mathfrak{F}}_1$, in agreement with our formula (\ref{lpbsv}) whose first line fits the SV model while the 2nd one is by itself a non propagating term. Moreover, the constraints on the sources for  ${\mathfrak{E}}_6$, see their figure 7, are in agreement with (\ref{sources}). The PBF and the Fronsdal models are not present in table II.

    In summary, we have an explicitly Lorentz invariant proof of unitarity of the new partially broken models at $s=3$ and $s=4$. The agreement of the amplitudes up to contact terms shows that the particle content is the same of their unbroken counterparts but indicates that there is no ``strong'' duality among those models such that a dual map would hold at action level where we have the product of fields at the same point, see comments in \cite{renato2} and also \cite{gracia}. So we do not expect full equivalence of possible interacting vertices in both theories. This is under investigation.

	\section{Acknowledgements}
	
	The work of D.D. is partially supported by CNPq  (grant 306380/2017-0) RSB is supported by CAPES while BSM is supported by FAPESP (2025/04628-1 ).

\section{Appendix: Projection operators on rank-3 and 4 fields} 
	
	We begin by introducing the standard projectors that decompose vector fields into longitudinal and transverse components, respectively:
	\be
	\omega_{\mu\nu} = \frac{\partial_\mu \partial_\nu}{\Box} 
	\quad ; \quad
	\theta_{\mu\nu} = \eta_{\mu\nu} - \omega_{\mu\nu}.
	\label{omegatheta}
	\ee
	These satisfy the algebraic identities and completeness relation:
	\be 
	\theta_{\mu\alpha} \theta^{\alpha\nu} = \theta_{\mu}^{\nu}, 
	\qquad 
	\omega_{\mu\alpha} \omega^{\alpha\nu} = \omega_{\mu}^{\nu}, 
	\qquad 
	\omega_{\mu\nu} \theta^{\nu\alpha} = 0, 
	\qquad 
	\delta^{\mu}_{\nu} = \theta^{\mu}_{\nu} + \omega^{\mu}_{\nu}. 
	\label{completeza}
	\ee
	
	From these vector projectors, we construct a complete basis of operators acting on totally symmetric rank-3 tensors. Following the approach of \cite{SchimidtElias}, we define:
	\bea
	(P^{(3)}_{11})^{\mu\nu\rho}_{\alpha\beta\gamma} &=& \theta^{(\mu}_{(\alpha} \theta^\nu_\beta \theta^{\rho)}_{\gamma)} - (P^{(1)}_{11})^{\mu\nu\rho}_{\alpha\beta\gamma}, \label{first} \\
	(P^{(2)}_{11})^{\mu\nu\rho}_{\alpha\beta\gamma} &=& 3\, \theta^{(\mu}_{(\alpha} \theta^\nu_\beta \omega^{\rho)}_{\gamma)} - (P^{(0)}_{11})^{\mu\nu\rho}_{\alpha\beta\gamma}, \\
	(P^{(1)}_{11})^{\mu\nu\rho}_{\alpha\beta\gamma} &=& \frac{3}{(D+1)} \theta^{(\mu\nu} \theta_{(\alpha\beta} \theta^{\rho)}_{\gamma)}, \\
	(P^{(1)}_{22})^{\mu\nu\rho}_{\alpha\beta\gamma} &=& 3\, \theta^{(\mu}_{(\alpha} \omega^\nu_\beta \omega^{\rho)}_{\gamma)}, \\
	(P^{(0)}_{11})^{\mu\nu\rho}_{\alpha\beta\gamma} &=& \frac{3}{(D-1)} \theta^{(\mu\nu} \theta_{(\alpha\beta} \omega^{\rho)}_{\gamma)}, \label{p110} \\
	(P^{(0)}_{22})^{\mu\nu\rho}_{\alpha\beta\gamma} &=& \omega_{\alpha\beta} \omega^{\mu\nu} \omega^{\rho}_{\gamma}. \label{p220}
	\eea
	
	The totally symmetric identity operator $\mathbbm{1}$ acting on rank-3 tensors is defined by:
	\be
	\mathbbm{1}^{\mu\nu\rho}_{\alpha\beta\gamma} = \delta^{(\mu}_{(\alpha} \delta^\nu_\beta \delta^{\rho)}_{\gamma)}. \label{id3}
	\ee
	
	\noindent Finally, the transition operators connecting different spin sectors are given by:
	\bea
	(P^{(1)}_{12})^{\mu\nu\rho}_{\alpha\beta\gamma} &=& \frac{3}{\sqrt{D+1}} \theta_{(\alpha\beta} \theta^{(\mu}_{\gamma)} \omega^{\nu\rho)}, \\
	(P^{(1)}_{21})^{\mu\nu\rho}_{\alpha\beta\gamma} &=& \frac{3}{\sqrt{D+1}} \theta^{(\mu\nu} \theta^{\rho)}_{(\alpha} \omega_{\beta\gamma)}, \\
	(P^{(0)}_{12})^{\mu\nu\rho}_{\alpha\beta\gamma} &=& \frac{3}{\sqrt{3(D-1)}} \theta_{(\alpha\beta} \omega^{(\mu\nu} \omega^{\rho)}_{\gamma)}, \\
	(P^{(0)}_{21})^{\mu\nu\rho}_{\alpha\beta\gamma} &=& \frac{3}{\sqrt{3(D-1)}} \theta^{(\mu\nu} \omega_{(\alpha\beta} \omega^{\rho)}_{\gamma)}.
	\eea
	
	Now we present a complete basis of spin projection and transition operators on totally symmetric rank-4 tensors in $D$ dimensions. The spin projectors acting on symmetric rank-4 tensors are given by:
	\begin{eqnarray}
		(P^{(4)}_{11})^{(\mu\nu\rho\sigma)}_{(\alpha\beta\gamma\lambda)} &=&   \theta^{(\mu}_{(\alpha} \theta^\nu_\beta \theta^\rho_\gamma \theta^{\sigma)}_{\lambda)} - (P^{(2)}_{11})^{(\mu\nu\rho\sigma)}_{(\alpha\beta\gamma\lambda)}-(P^{(0)}_{11})^{(\mu\nu\rho\sigma)}_{(\alpha\beta\gamma\lambda)}, \\		
		(P^{(2)}_{11})^{(\mu\nu\rho\sigma)}_{(\alpha\beta\gamma\lambda)} &=&\frac{6}{(D+3)}\,\,\left\lbrack \theta_{(\alpha\beta}\theta^{(\mu\nu}\theta^\rho_\gamma \theta^{\sigma)}_{\lambda)} -\frac{(D+1)}{3}\,\,(P^{(0)}_{11})^{(\mu\nu\rho\sigma)}_{(\alpha\beta\gamma\lambda)}\right\rbrack\\ 
		(P^{(0)}_{11})^{(\mu\nu\rho\sigma)}_{(\alpha\beta\gamma\lambda)} &=& \frac{3}{(D-1)(D+1)} \,\, \theta^{(\mu\nu}\theta^{\rho\sigma)} \theta_{(\alpha\beta} \theta_{\gamma\lambda)},  \\
		(P^{(3)}_{11})^{(\mu\nu\rho\sigma)}_{(\alpha\beta\gamma\lambda)} &=& 4\,\,  \theta^{(\mu}_{(\alpha} \theta^\nu_\beta \theta^\rho_\gamma \omega^{\sigma)}_{\lambda)}-(P^{(1)}_{11})^{(\mu\nu\rho\sigma)}_{(\alpha\beta\gamma\lambda)},  \\
		(P^{(1)}_{11})^{(\mu\nu\rho\sigma)}_{(\alpha\beta\gamma\lambda)} &=& \frac{12}{(D+1)}\theta_{(\alpha\beta}\theta^{(\mu\nu}\theta^\rho_\gamma \omega^{\sigma)}_{\lambda)} \\
		(P^{(2)}_{22})^{(\mu\nu\rho\sigma)}_{(\alpha\beta\gamma\lambda)} &=& 6\,\,\theta^{(\mu}_{(\alpha} \theta^\nu_\beta \omega^\rho_\gamma \omega^{\sigma)}_{\lambda)}-(P^{(0)}_{22})^{(\mu\nu\rho\sigma)}_{(\alpha\beta\gamma\lambda)},\\
		(P^{(0)}_{22})^{(\mu\nu\rho\sigma)}_{(\alpha\beta\gamma\lambda)} &=& \frac{6}{(D-1)}\,\,\theta^{(\mu\nu} \theta_{\alpha\beta} \, \omega^\rho_\gamma \omega^{\sigma)}_{\lambda)},\\
		(P^{(1)}_{22})^{(\mu\nu\rho\sigma)}_{(\alpha\beta\gamma\lambda)} &=& 4\,\,\theta^{(\mu}_{(\alpha} \omega^\nu_\beta \omega^\rho_\gamma \omega^{\sigma)}_{\lambda)},\\
		(P^{(0)}_{33})^{(\mu\nu\rho\sigma)}_{(\alpha\beta\gamma\lambda)} &=& \,\,\omega^{(\mu}_{(\alpha} \omega^\nu_\beta \omega^\rho_\gamma \omega^{\sigma)}_{\lambda)},. 
	\end{eqnarray}
	with the symmetric rank-4 identity operator given by:
	\begin{equation}
		\mathbbm{1}^{(\mu\nu\rho\sigma)}_{(\alpha\beta\gamma\lambda)} = \delta^{(\mu}_{(\alpha} \delta^\nu_\beta \delta^\rho_\gamma \delta^{\sigma)}_{\lambda)}.
	\end{equation}
	
	The corresponding transition operators for $s = 0, 1, 2$ sectors are:
	\begin{eqnarray}
		(P^{(0)}_{{12}})^{(\mu\nu\rho\sigma)}_{(\alpha\beta\gamma\lambda)} &=& \frac{6}{(D-1)\sqrt{2(D+1)}}   \theta^{(\mu\nu}\omega^{\rho\sigma)}\theta_{(\alpha\beta}\theta_{\gamma\lambda)}  \\ 
		(P^{(0)}_{{21}})^{(\mu\nu\rho\sigma)}_{(\alpha\beta\gamma\lambda)} &=& \frac{6}{(D-1)\sqrt{2(D+1)}}   \theta^{(\mu\nu}\theta^{\rho\sigma)}\theta_{(\alpha\beta}\omega_{\gamma\lambda)},  \\
		(P^{(0)}_{{13}})^{(\mu\nu\rho\sigma)}_{(\alpha\beta\gamma\lambda)} &=&  \frac{3}{\sqrt{3(D-1)(D+1)}}   \omega^{(\mu\nu}\omega^{\rho\sigma)}\theta_{(\alpha\beta}\theta_{\gamma\lambda)},   \\ 
		(P^{(0)}_{{31}})^{(\mu\nu\rho\sigma)}_{(\alpha\beta\gamma\lambda)} &=& \frac{3}{\sqrt{3(D-1)(D+1)}}   \theta^{(\mu\nu}\theta^{\rho\sigma)}\omega_{(\alpha\beta}\omega_{\gamma\lambda)},\\
		(P^{(0)}_{{23}})^{(\mu\nu\rho\sigma)}_{(\alpha\beta\gamma\lambda)} &=& \frac{6}{\sqrt{6(D-1)}}   \omega^{(\mu\nu}\omega^{\rho\sigma)}\theta_{(\alpha\beta}\omega_{\gamma\lambda)},\\
		(P^{(0)}_{{32}})^{(\mu\nu\rho\sigma)}_{(\alpha\beta\gamma\lambda)} &=& \frac{6}{\sqrt{6(D-1)}}   \theta^{(\mu\nu}\omega^{\rho\sigma)}\omega_{(\alpha\beta}\omega_{\gamma\lambda)}\\
		(P^{(1)}_{{12}})^{(\mu\nu\rho\sigma)}_{(\alpha\beta\gamma\lambda)} &=& \frac{12}{\sqrt{3(D+1)}}   \theta_{(\alpha\beta}\theta^{(\mu}_{\gamma}\omega^{\nu}_{\lambda)}\omega^{\rho\sigma)},\\
		(P^{(1)}_{{21}})^{(\mu\nu\rho\sigma)}_{(\alpha\beta\gamma\lambda)} &=& \frac{12}{\sqrt{3(D+1)}}   \theta^{(\mu\nu}\theta^{\rho}_{(\alpha}\omega^{\sigma)}_{\beta}\omega_{\gamma\lambda)}\\
		(P^{(2)}_{{12}})^{(\mu\nu\rho\sigma)}_{(\alpha\beta\gamma\lambda)} &=& \frac{6}{\sqrt{(D+3)}}   \left\lbrack \theta_{(\alpha\beta}\theta^{(\mu}_{\gamma}\theta^{\nu}_{\lambda)}\omega^{\rho\sigma)} -\frac{\sqrt{2(D+1)}}{6}( P^{(0)}_{{12}})^{(\mu\nu\rho\sigma)}_{(\alpha\beta\gamma\lambda)}\right\rbrack,\\
		(P^{(2)}_{{21}})^{(\mu\nu\rho\sigma)}_{(\alpha\beta\gamma\lambda)} &=& \frac{6}{\sqrt{(D+3)}}      \left\lbrack \theta^{(\mu\nu}\theta^{\rho}_{(\alpha}\theta^{\sigma)}_{\beta}\omega_{\gamma\lambda)} -\frac{\sqrt{2(D+1)}}{6}( P^{(0)}_{{21}})^{(\mu\nu\rho\sigma)}_{(\alpha\beta\gamma\lambda)}\right\rbrack. 
	\end{eqnarray}
	
	These operators in both  cases (rank-3 and rank-4) satisfy the algebra:
	\begin{equation}
		P_{ij}^{(s)} P_{kl}^{(r)} = \delta^{sr} \delta_{jk} P_{il}^{(s)}. \label{algebra}
	\end{equation}
	
	Here, superscripts $(r)$ and $(s)$ denote the spin sector, while the subscripts label different operators. If $i = j$ (or $k = l$), the operator is a projector; otherwise, it is a transition operator. The projectors resolve the identity:
	\begin{equation}
		\sum_{i,s} P^{(s)}_{ii} = \mathbbm{1},
	\end{equation}
	The transition operators satisfy the algebra in (\ref{algebra}), but are not required for the identity decomposition. Still, they are crucial for expanding bilinear structures involving two symmetric rank-3 or 4 fields in a Lagrangian.
	
	In the expressions above, parenthesis denote normalized symmetrization. For instance, the first term in Eq.~\eqref{first} expands as:
	\be
	\theta^{(\mu}_{(\alpha} \theta^\nu_\beta \theta^{\rho)}_{\gamma)} = \frac{1}{6} \left(
	\theta^{\mu}_{\alpha} \theta^\nu_\beta \theta^{\rho}_{\gamma} +
	\theta^{\rho}_{\alpha} \theta^\nu_\beta \theta^{\mu}_{\gamma} +
	\theta^{\nu}_{\alpha} \theta^\mu_\beta \theta^{\rho}_{\gamma} +
	\theta^{\rho}_{\alpha} \theta^\mu_\beta \theta^{\nu}_{\gamma} +
	\theta^{\nu}_{\alpha} \theta^\rho_\beta \theta^{\mu}_{\gamma} +
	\theta^{\mu}_{\alpha} \theta^\rho_\beta \theta^{\nu}_{\gamma} \right).
	\ee


\begin{thebibliography}{99}
	
	\bibitem{fp}
	M.~Fierz,
	``On the relativistic theory of free particles with any spin,''
	\textit{Helv. Phys. Acta} \textbf{12}, 3 (1939);
	M.~Fierz and W.~Pauli,
	``On relativistic wave equations for particles of arbitrary spin in an electromagnetic field,''
	\textit{Proc. R. Soc. Lond. A} \textbf{173}, 211 (1939).
	
	\bibitem{sh}
	L.~P.~S.~Singh and C.~R.~Hagen,
	``Lagrangian formulation for arbitrary spin. I. The boson case,''
	\textit{Phys. Rev. D} \textbf{9}, 4 (1974).
	
	\bibitem{fronsdal}
	C.~Fronsdal,
	``Massless Fields with Integer Spin,''
	\textit{Phys. Rev. D} \textbf{18}, 3624 (1978),
	doi:10.1103/PhysRevD.18.3624.
	
	\bibitem{fs1}
	D.~Francia and A.~Sagnotti,
	``Free geometric equations for higher spins,''
	\textit{Phys. Lett. B} \textbf{543}, 303 (2002),
	[arXiv:hep-th/0207002].
	
	\bibitem{vasiliev_r}
	X.~Bekaert, S.~Cnockaert, C.~Iazeolla and M.~A.~Vasiliev,
	``Nonlinear higher-spin theories in various dimensions,''
	[arXiv:hep-th/0503128].
	
	\bibitem{st_r}
	A.~Sagnotti and M.~Taronna,
	``String Lessons for Higher-Spin Interactions,''
	\textit{Nucl. Phys. B} \textbf{842}, 299–361 (2011),
	[arXiv:1006.5242 [hep-th]].
	
	\bibitem{rt}
	R.~Rahman and M.~Taronna,
	``From Higher Spins to Strings: A Primer,''
	[arXiv:1512.07932 [hep-th]].
	
	\bibitem{snow}
	X.~Bekaert, N.~Boulanger, A.~Campoleoni, M.~Chiodaroli, D.~Francia, M.~Grigoriev, E.~Sezgin and E.~Skvortsov,
	``Snowmass White Paper: Higher Spin Gravity and Higher Spin Symmetry,''
	[arXiv:2205.01567 [hep-th]].
	
	\bibitem{book}
	A.~Bengtsson,
	\textit{Higher Spin Field Theory, Vol. 1+2},
	De Gruyter, 2023.
	
	\bibitem{vasilievr2}
	M.~A.~Vasiliev,
	\textit{JHEP} \textbf{07}, 110 (2025),
	e-Print:2503.10967 [hep-th].
	
	\bibitem{blas}
	E.~Alvarez, D.~Blas, J.~Garriga and E.~Verdaguer,
	``Transverse Fierz–Pauli Symmetry,''
	\textit{Nucl. Phys. B} \textbf{756}, 148 (2006);
	D.~Blas,
	``Aspects of Infrared Modifications of Gravity,''
	PhD Thesis, University of Barcelona,
	[arXiv:0809.3744].
	
	\bibitem{sv}
	E.~Skvortsov and M.~Vasiliev,
	``Transverse Invariant Higher Spin Fields,''
	\textit{Phys. Lett. B} \textbf{664}, 301 (2008),
	[arXiv:0701278 [hep-th]].
	
\bibitem{pbf}
	D.~Dalmazi and B.~dos~S.~Martins,
	``A partially broken Fronsdal model for massless higher-spin particles of integer spin,'', JHEP 10 (2025) 126, e-Print:2507.04226.	
	
	\bibitem{tdiffspin3}
	R.~S.~Bittencourt, D.~Dalmazi, B.~dos~S.~Martins and E.~L.~Mendonça,
	``Equivalence of spin-2 and spin-3 models invariant under transverse diffeomorphisms and the tensionless limit of string theory,''
	\textit{Phys. Rev. D} \textbf{112}, 025008 (2025),
	doi:10.1103/jvj7-3m11,
	[arXiv:2501.04214 [hep-th]].
	
	\bibitem{renato2}
	D.~Dalmazi and R.~C.~Santos,
	``Note on linearized new massive gravity in arbitrary dimensions''
	\textit{Phys. Rev. D} \textbf{87}, 085021 (2013), e-Print:1212.6753 [hep-th].
	
	\bibitem{gracia}
	G.~B.~de~Gracia and B.~M.~Pimentel,
	\textit{Eur. Phys. J. Plus} \textbf{139}, 248 (2024),
	e-Print:2403.10793 [hep-th].
	
	\bibitem{bms3}
	W.~Barker, C.~Marzo and A.~Santoni,
	``Infrared foundations for quantum geometry I: Catalogue of totally symmetric rank-three field theories,''
	e-Print:2506.21662 [hep-th].
	
	\bibitem{SchimidtElias}
	E.~L.~Mendonça and R.~Schimidt~Bittencourt,
	``Unitarity of Singh-Hagen Model in D Dimensions,''
	\textit{Adv. High Energy Phys.} \textbf{2020}, 8425745 (2020),
	doi:10.1155/2020/8425745,
	[arXiv:1902.05118 [hep-th]].
	
\end{thebibliography}
\end{document}